\documentclass[times]{elsarticle}

\usepackage{bsymb}

\usepackage{cite}
\usepackage{amsmath,amssymb,amsfonts}
\usepackage{algorithmic}
\usepackage{graphicx}
\usepackage{subfig}

\usepackage{textcomp}
\def\BibTeX{{\rm B\kern-.05em{\sc i\kern-.025em b}\kern-.08em
    T\kern-.1667em\lower.7ex\hbox{E}\kern-.125emX}}

\usepackage{mathtools}

\usepackage[utf8]{inputenc}
\usepackage[T1]{fontenc}
\usepackage{microtype}

\usepackage{color}
\definecolor{gray}{rgb}{0.4,0.4,0.4}
\definecolor{darkblue}{rgb}{0.0,0.0,0.6}
\definecolor{cyan}{rgb}{0.0,0.6,0.6}
\definecolor{keycolor}{rgb}{0,0,0.8}     
\definecolor{labelcolor}{rgb}{0,0.4,0.8} 
\definecolor{codecolor}{rgb}{0,0,0}      
\definecolor{inhcolor}{rgb}{0.6,0.2,0}   
\definecolor{cmtcolor}{rgb}{0,0.4,0}     

\usepackage{bsymb}

\usepackage{listings}

\usepackage{color}
\definecolor{gray}{rgb}{0.4,0.4,0.4}
\definecolor{darkblue}{rgb}{0.0,0.0,1.0}
\definecolor{cyan}{rgb}{0.0,0.6,0.6}

\lstdefinelanguage{MPS}
{
  morestring=[b]",
  morestring=[s]
  ,
  morecomment=[s]{/*}{*/},
  stringstyle=\color{black},
  identifierstyle=\color{black},
  keywordstyle=\color{darkblue},
  morekeywords={domain,model,parent,concepts,concept,is,variable,
  individuals,deduced,attribute,domain,dom,relations,relation,attributes,
  enumerated,elements,
  range,functional,total,maplets,custom,data,sets,set,values,value,type,lexical,form,
  predicates,p1,p2,not}
  ,
  otherkeywords = {=,&,(,),\{,\},>,<,",:,?},
}

\usepackage{lineno,hyperref}
\modulolinenumbers[5]

\journal{Arxiv}

\usepackage{breqn}
\usepackage{longtable}

\usepackage[linewidth=1pt]{mdframed}
\usepackage{multicol}

\usepackage{flushend}

\begin{document}

\begin{frontmatter}

\title{Pear2Pear (On Wifi): \\A Data Sharing Protocol Over Wifi through a Peer to Peer Network
}


\author{Steve Tueno, Romeo Tabue, Forentin Jiechieu, Yacynth Ndonna, Billy Zafack, Audric Feuyan, Jonas Atibita, Alex Djouontse and  Rodrigue  Mbinkeu}
\address{National Advanced School of Engineering, Yaoundé, Cameroon}
\ead{stuenofotso@gmail.com, steve.jeffrey.tueno.fotso@usherbrooke.ca}


\begin{abstract}
A peer-to-peer system is a distributed system in which equal nodes (in terms of role and usage) exchange information and services directly.
This paper describes a distributed peer-to-peer protocol that allows wifi-enabled smart devices (especially Android smartphones) to exchange data using only wifi. The protocol is designed to allow the automatic establishment of a distributed peer-to-peer network of any size without geography constraint.
It is applied to file sharing between devices, but can  easily be adapted to support any data sharing.
The protocol defines two layers: (1) a \textit{kernel layer} responsible for creating, routing, establishing and maintaining links between nodes (peer), addressing a node and adding and removing nodes; and an (2) application layer to support data sharing (file sharing in this case).
The structure of the peer-to-peer network  is hybrid. 
Regarding the file sharing use case, the application allows a node to (i) search for a file in the catalog of its subnet that is held by the \textit{root} node, and (ii) download a file: if the file is not in the subnet, the root node delegates a node to make vouchers in neighboring subnetworks to get the file and make it available.
A proof of concept was made on Android.

\end{abstract}


\end{frontmatter}


\section*{Introduction}
Since the beginning of the Internet, the client-server model was the reference model for the provision of resources. In this model, the system relies on a dedicated server that centralizes and maintains all resources and services. Therefore, the increase in the number of users requires a greater investment of service providers. It is indeed required to guarantee the availability of resources and services, despite the large number of simultaneous requests. This requires significant resources and imposes software and hardware constraints, which make such systems very expensive and difficult to secure and scale because of the single point of failure. However, the increase in the number of users also implies an increase in cumulative resources (computing and storage), hence the  peer-to-peer paradigm.

The peer-to-peer paradigm consists of connecting users in order to pool resources and distribute tasks \citep{DBLP:conf/vldb/YangG01}. A peer-to-peer system is thus a distributed system of connected peers, consumers and service providers. Systems based on this paradigm are implemented as logical networks, connecting participants over physical networks.
The work presented in this paper reports on the design of an hybrid peer-to-peer protocol in which peers are wifi-enabled mobiles not connected to Internet.

\section{Background}

\subsection{The Peer-to-Peer Paradigm} 
The peer-to-peer paradigm (P2P)  has been focused from the outset on the management of large amounts of data spread over a very large scale. Although it has been used so far mainly for file sharing purposes, many research projects are  interested in other types of applications including collaboration, distributed computing, sharing and distribution of content.

\subsubsection{Definition}
A peer-to-peer system is a distributed system in which equal nodes (in terms of role and usage) directly  exchange information and services \citep{DBLP:conf/vldb/YangG01}. Behind this academic definition is the reality of the context of use of these systems. Indeed, it is the role of peers to put and maintain each peer-to-peer system  in place. There is  no central entity that controls the system. This entails a number of advantages, but also constraints. For benefits, increasing the number of participants also increases the resources available while distributing the load among participants. It is therefore theoretically possible to scale to a relatively low cost. In addition, the absence of a central control entity gives the system greater robustness, since the availability of resources is not directly related to a particular peer \citep{dalle2008analysis}. For constraints, the absence of a central entity requires the establishment of a self-organizing mechanism. Peers must connect to or disconnect from the network without affecting the availability of resources. In addition, peer-to-peer systems must manage a network of highly heterogeneous nodes (hardware architecture: processor, memory and connection type: network interface, connection rate, ...).

\subsubsection{Different centralization levels}
\begin{figure}[!h]
\begin{center}
\includegraphics[width=.52\textwidth]{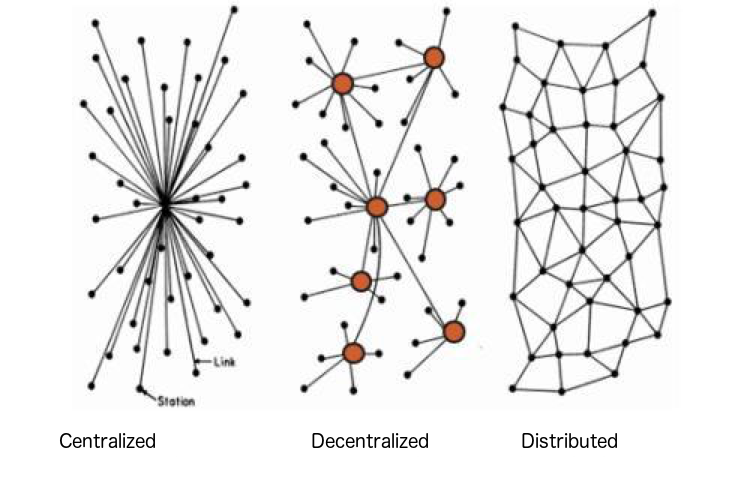}
\end{center}
\caption{\label{p2p_systems_classification} P2P systems classification}
\end{figure}

P2P systems can be classified into three major families according to the level of centralization that results in a more or less important distribution of tasks to be accomplished. These are centralized, distributed and decentralized (hybrid) P2P models (Fig. \ref{p2p_systems_classification}).

The \textit{centralized model} is at the boder of the P2P model, because it is based on a dedicated server that centralizes and maintains the body of knowledge of peers, the resources always being hosted on peers. In this model, the server is a location entity and contains no resources. Only the retrieval of resources is decentralized. The advantage of this model is that it allows a comprehensive search, fast resource localization and simple management thanks to the central entity.
The major disadvantage is that the system operation is based solely on the central entity. The latter must be able to support a large number of queries and perform a lot of research.
A well-known example falling into this category is that of the  file-sharing system \textit{MP3 Napster} \citep{DBLP:journals/mms/SaroiuGG03}.

\textit{Distributed P2P systems} try to distribute all the system functions among the peers, namely: searching, routing and retrieving objects in the logical network. Peers need to organize to form an effective dynamic architecture connecting them to each other.
There are two approaches to this: build a random graph and rely on random explorations/searches of the network, or build a structured graph in which an efficient routing is gradually supported \citep{viennot2005autour}.
A well-known example falling into this category is that of   \textit{Gnutella v0.4} \citep{ripeanu2001peer}. Gnutella in its first version v0.4, proposes to connect the nodes in a random way and uses the flood for its searches in the logical network thus formed. For a better use of resources, the following version, \textit{Gnutella v0.6}, proposes to organize the logical network according to a decentralized architecture on two levels.

The \textit{decentralized model}, also called hybrid model, adds an  hierarchy to the centralized model. This model relies on super node interconnections at the top of the hierarchy, according to the distributed model. Each leaf node is attached to one or more super nodes. A super node manages a set of leaf nodes. Objects shared by a leaf node are saved to the super node responsible for that node. When a leaf node looks for an object, it sends its request to its super node. This one then carries out the search among the objects of the nodes which are attached to it, or even the neighboring super nodes if necessary.
A well-known example falling into this category is that of   \textit{FastTrack} \citep{liang2006fasttrack}: 1 super node for 100 leaf nodes.

Existing peer-to-peer systems rely on  Internet for data sharing. Our goal is to design a peer-to-peer system able to circumvent the need for an active Internet connexion by relying solely on  device wifi network interfaces.

\section{The Pear2Pear On Wifi Protocol}
The protocol is designed according to the decentralized or hybrid model.

\begin{figure}[!h]
\begin{center}
\includegraphics[width=1.0\textwidth]{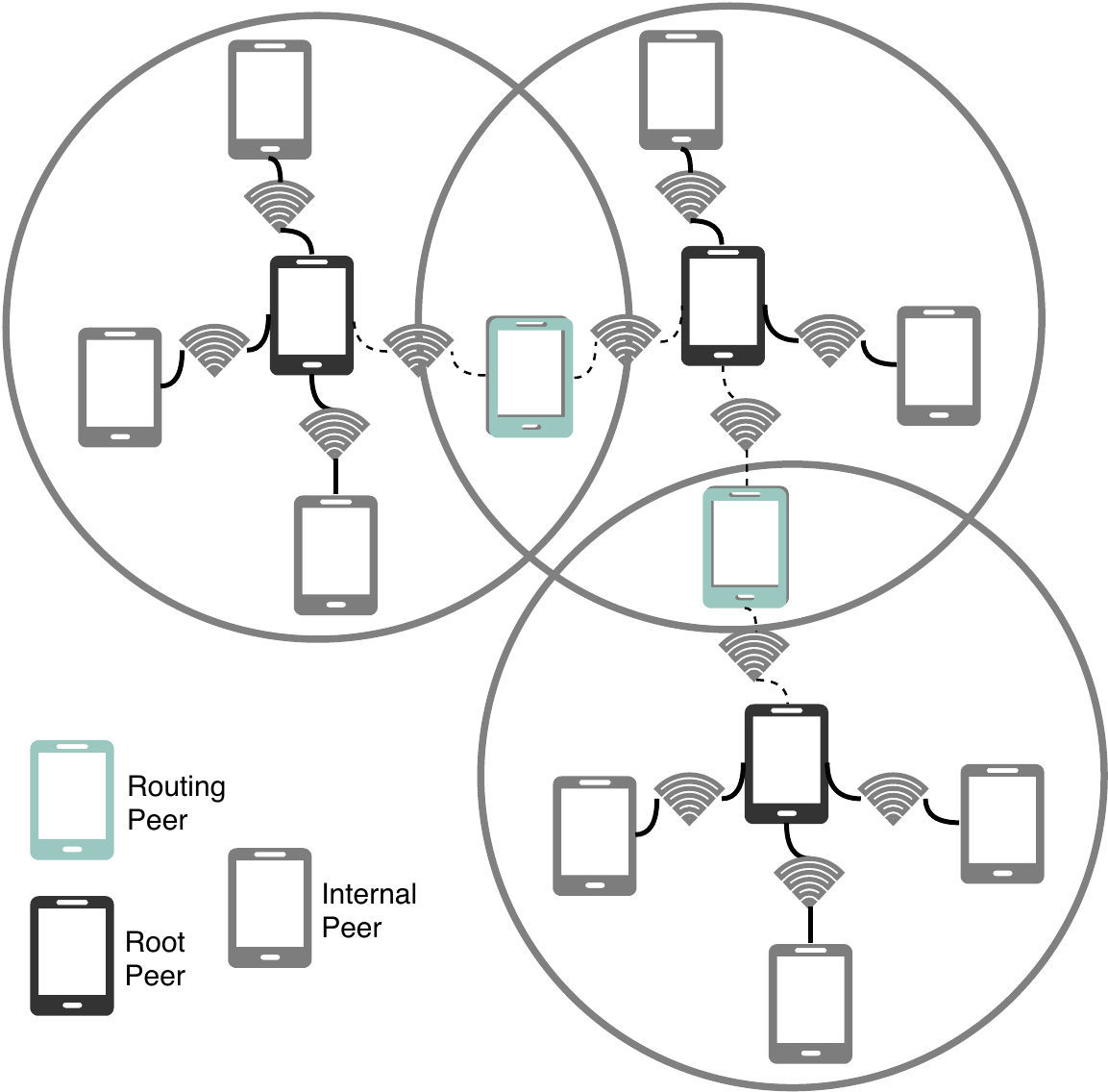}
\end{center}
\caption{\label{Pear2Pear_architectural_view} Architectural View of  Pear2Pear On Wifi Subnetworks}
\end{figure}

\begin{figure}[!h]
\begin{center}
\includegraphics[width=.7\textwidth]{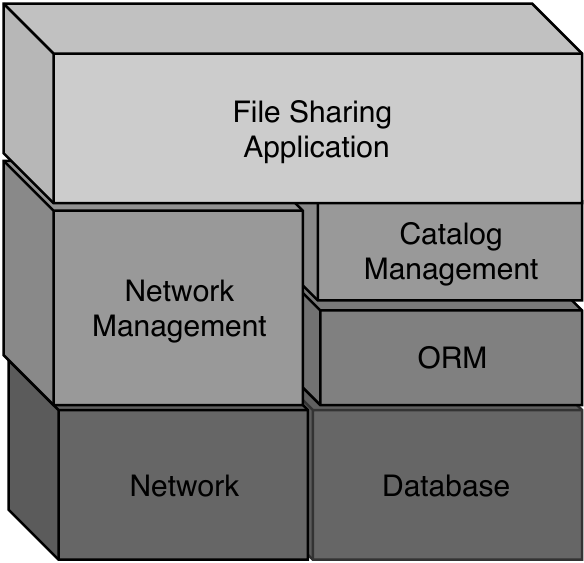}
\end{center}
\caption{\label{pear2pear_protocol} Building Blocks of  a Pear2Pear On Wifi Subnetwork}
\end{figure}

\begin{figure}
  \subfloat[]{%
  \begin{minipage}{.5\linewidth}
  \includegraphics[width=.8\linewidth]{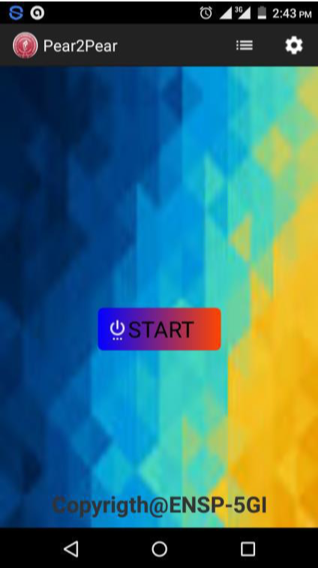}
  \end{minipage}%
  }
  \subfloat[]{%
  \begin{minipage}{.5\linewidth}
  \includegraphics[width=.8\linewidth]{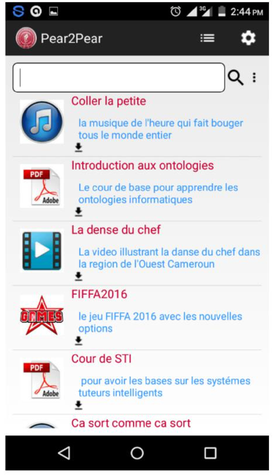}
  \end{minipage}%
  }
\caption{\label{apercu} Overview of the  Pear2Pear On Wifi Android Proof of Concept}
\end{figure}

A proof of concept was made on Android (Fig. \ref{apercu}).

\subsection{Problems / Contributions}

Several problems were identified as soon as this project started:
\begin{itemize}
\item[•] Establishing and maintaining links between nodes
\item[•] Adding and removing a node from the network (does a node exit the network or is it removed ?)
\item[•]  Routing (how to choose the best route for a given packet ?)
\item[•] Node addressing
\item[•] Management of available resources (battery, memory,...)
\item[•] File sharing protocol
\begin{itemize}
\item[•] How to identify a file on the network ? 
\item[•] Uniqueness, completeness, error handling for a given file
\item[•] Managing duplicate files (if a peer wants to download a file it already has)
\item[•] Storing files (file block, complete file)
\item[•] Parallelism management in file sharing
\item[•] Management of identical files but with different names
\end{itemize}

\end{itemize}
 
\subsection{Algorithms}
To solve the identified problems, several algorithms have been set up. This section presents some among  the most relevant ones.

\subsubsection{Subnet Creation}
A wifi-enabled device wants to connect:
\begin{enumerate}
\item It scans surrounding wifi networks for available networks.
\begin{itemize}
\item If it identifies a network, it connects to it; otherwise, it creates a network (hotspot) and becomes its \textit{root}. The ssid of each network is defined so as to make it unique and quickly identifiable by  other peers. Each network is protected. To access it, a complex computation function allows each device equipped with \textit{Pear2Pear} to compute the passphrase from the ssid.
\item  If it identifies a network and cannot connect to it after a while, it creates a network (hotspot) and becomes its \textit{root}. Indeed, some operating systems, including Android, limit the number of devices likely to join the hotspot. This number can also be limited by the protocol in order to optimize the use of ressources of the network root (battery, memory, ...).
\end{itemize}
\end{enumerate}

Regarding the management of peer  departures, if a peer notifies its root  of its departure, it is removed from  catalogs after  the expiration of a certain countdown. 
Moreover, the root must make pings to scan for available peers and deals with silent leaves.

\subsubsection{The Network File Catalog}
Each subnet is associated with a file catalog. The catalog identifies for each file the proprietary device and its subnet. Each file is represented  by its original name and an unique identifier, related to its content, obtained through the computation of the hash of this content. This makes it possible to manage the duplication of files and to optimize the acquisition of data.

\begin{enumerate}
\item The root device initializes the catalog of its subnet, when it is created, with the list of its shared files.
\item Each peer that connects to it, communicates the list of its available files.
\end{enumerate}

Each non-root device in a subnet periodically scans other accessible subnets and informs its root to allow it to build a subnet catalog. The subnet catalog identifies other accessible subnets and for each,  the list of peers that can access it.

\subsubsection{File Search and File Management}
\textbf{Note:} when a file search fails, a special frame must (or can) be emitted (once or periodically) in order to invite the peers to make this file available as far as possible.

To allow inter-network searches, each root must periodically designate, for each neighboring subnet, a peer that will switch to that subnet and retrieve the file catalog to bring it back to its original subnet and make it available to the root.  This will enrich the network file catalog of the subnet which will  support  all file requests. A subnet that is no longer accessible is automatically removed from the catalog after the expiration of a certain countdown. Each time a file is added or deleted, a notification is sent to the root to update the catalog.

The choice of the peer which will retrieve the catalog of another subnet, among all peers having access to this subnet, must be done by an algorithm ensuring optimal use of peer resources. We choosed \textit{round robin} for our proof of concept.

A catalog makes it possible to reference for each subnet whose file is referenced the number of jumps required to reach it and  the nearest network that serves as  gateway to it. This supports the computation of the shortest path  to retrieve a file (also taking into account file duplicates that enable distribute file retrieval).

\subsubsection{File Retrieval}
A peer that searches for an available file gets the unique identifier of the file.

\paragraph{A Peer to a Peer of the Same Subnet}:
A peer \textit{P1} of the subnet \textit{x} wants to retrieve the file \textit{f} present in the shared folder of peer \textit{P2}  of the subnet \textit{x}:

\begin{enumerate}
\item \textit{P1} sends a request to download the file \textit{f} specifying the unique identifier of \textit{f} in the network file catalog.
\item The root of subnet  \textit{x} identifies the peer \textit{P2} as the closest (same subnet) peer that has a copy  of the requested file and
 sends to \textit{P1} the network address of \textit{P2}.
 \item \textit{P1} proceeds directly to download the file from \textit{P2}.
\end{enumerate}

If multiple peers of \textit{x} have a copy of the file, the root makes all their addresses available to \textit{P1} so that it can proceeds with a distributed download of the blocks of \textit{f}.

\paragraph{A Peer to a Peer of a Different Subnet}:
The peer \textit{P1} of the subnet \textit{S1} wants to retrieve the file \textit{f} of a peer  of the subnet \textit{Sn}, where \textit{n} is the number of hops separating the 2 subnets:

\begin{enumerate}
\item \textit{P1} sends a request to download the file \textit{f} specifying the unique identifier of \textit{f} in the network file catalog.

\item The root of \textit{S1} identifies the closest subnet that owns the file,  based on available file copies and  number of hops between networks, as \textit{Sn}. It also identifies the peer \textit{P1'} of \textit{S1}  as the most likely to go to \textit{S2}, the gateway from \textit{S1} to \textit{Sn}. Therefore,  it sends \textit{P1'} to \textit{S2} with the file request.

\item \textit{P1'}, once in \textit{S2}, checks if $S2 == Sn$. 

\item If $S2 == Sn$, it sends the file request to the root of \textit{S2} to receive the network address of the peer \textit{P2} of \textit{S2}  hosting the file.

 \item \textit{P1'} proceeds directly to download the file from \textit{P2}.
 
 \item If $S2 /= Sn$, the protocol resumes in step 1  with $P1=P1'$ and $S1 = S2$.
 
 \item \textit{P1'} jumps back to  \textit{S1} and directly delivers the file  to \textit{P1} since it has its  network address.

\end{enumerate}

If multiple peers of \textit{S2} have a copy of the file, the root makes all their addresses available to \textit{P1'} so that it can proceeds with a distributed download of the blocks of \textit{f}.
In addition, if the file is very large, several peers of \textit{S1} having access to \textit{S2} can jump to \textit{S2} to request specific blocks of \textit{f} and return to distributively deliver it  to \textit{P1}.

\section*{References}



\bibliographystyle{elsarticle-num}
\bibliography{references}

\begin{thebibliography}{1}
\expandafter\ifx\csname url\endcsname\relax
  \def\url#1{\texttt{#1}}\fi
\expandafter\ifx\csname urlprefix\endcsname\relax\def\urlprefix{URL }\fi
\expandafter\ifx\csname href\endcsname\relax
  \def\href#1#2{#2} \def\path#1{#1}\fi

\bibitem{DBLP:conf/vldb/YangG01}
B.~Yang, H.~Garcia{-}Molina,
  \href{http://www.vldb.org/conf/2001/P561.pdf}{Comparing hybrid peer-to-peer
  systems}, in: P.~M.~G. Apers, P.~Atzeni, S.~Ceri, S.~Paraboschi,
  K.~Ramamohanarao, R.~T. Snodgrass (Eds.), {VLDB} 2001, Proceedings of 27th
  International Conference on Very Large Data Bases, September 11-14, 2001,
  Roma, Italy, Morgan Kaufmann, 2001, pp. 561--570.
\newline\urlprefix\url{http://www.vldb.org/conf/2001/P561.pdf}

\bibitem{dalle2008analysis}
O.~Dalle, F.~Giroire, J.~Monteiro, S.~P{\'e}rennes, Analysis of failure
  correlation in peer-to-peer storage systems, Ph.D. thesis, INRIA (2008).

\bibitem{DBLP:journals/mms/SaroiuGG03}
S.~Saroiu, P.~K. Gummadi, S.~D. Gribble,
  \href{https://doi.org/10.1007/s00530-003-0088-1}{Measuring and analyzing the
  characteristics of napster and gnutella hosts}, Multimedia Syst. 9~(2) (2003)
  170--184.
\newblock \href {http://dx.doi.org/10.1007/s00530-003-0088-1}
  {\path{doi:10.1007/s00530-003-0088-1}}.
\newline\urlprefix\url{https://doi.org/10.1007/s00530-003-0088-1}

\bibitem{viennot2005autour}
L.~Viennot, Autour des graphes et du routage, Ph.D. thesis, Universit{\'e}
  Paris-Diderot-Paris VII (2005).

\bibitem{ripeanu2001peer}
M.~Ripeanu, Peer-to-peer architecture case study: Gnutella network, in:
  Proceedings first international conference on peer-to-peer computing, IEEE,
  2001, pp. 99--100.

\bibitem{liang2006fasttrack}
J.~Liang, R.~Kumar, K.~W. Ross, The fasttrack overlay: A measurement study,
  Computer Networks 50~(6) (2006) 842--858.

\end{thebibliography}

\end{document}